\documentclass[conference]{IEEEtran}
\usepackage{cite}

\ifCLASSINFOpdf
   \usepackage[pdftex]{graphicx}
  \graphicspath{{../pdf/}{../jpeg/}}
  \DeclareGraphicsExtensions{.pdf,.jpeg,.png}
\else
  \usepackage[dvips]{graphicx}
  \graphicspath{{../eps/}}
   \DeclareGraphicsExtensions{.eps}
\fi

\usepackage[cmex10]{amsmath}
\usepackage{rotating}
\usepackage{amsmath,amsthm,verbatim,amssymb,amsfonts,amscd, graphicx}
\usepackage{array}
\usepackage{multicol,multirow}

\usepackage{algorithmic}
\usepackage{algorithm}
\usepackage{citesort}

\usepackage{array}
\usepackage{balance}

\ifCLASSOPTIONcompsoc
\usepackage[caption=false,font=normalsize,labelfont=sf,textfont=sf]{subfig}
\else
\usepackage[caption=false,font=footnotesize]{subfig}
\usepackage{color}

\usepackage{url}


\def\maketag@@@#1{\hbox{\m@th\normalfont\normalsize#1}}

\begin{document}
\title{Coordinated Two-Tier Heterogeneous Cellular Networks with Leakage Based Beamforming}

\author{\IEEEauthorblockN{Harsh Tataria$^{*}$, Mansoor Shafi$^{\dagger}$, Peter J. Smith$^{\ddagger}$ and Pawel A. Dmochowski$^{*}$}
\IEEEauthorblockA{$^{*}$School of Engineering and Computer Science, Victoria University of Wellington, Wellington, New Zealand.\\$^{\dagger}$Spark New Zealand, Wellington, New Zealand.\\$^{\ddagger}$ Department of Electrical and Computer Engineering, University of Canterbury, Christchurch, New Zealand.}email:\{Harsh.Tataria, Pawel.Dmochowski\}@ecs.vuw.ac.nz, Mansoor.Shafi@spark.co.nz, Peter.Smith@canterbury.ac.nz}
\maketitle

\begin{abstract}
In this paper we demonstrate the rate gains achieved by two-tier heterogeneous cellular networks (HetNets) with varying degrees of coordination between macrocell and microcell base stations (BSs). We show that without the presence of coordination, network densification does not provide any gain in the sum rate and rapidly decreases the mean per-user signal-to-interference-plus-noise-ratio (SINR). Our results show that coordination reduces the rate of SINR decay with increasing numbers of microcell BSs in the system. Validity of the analytically approximated mean per-user SINR over a wide range of signal-to-noise-ratio (SNR) is demonstrated via comparison with the simulated results.
\end{abstract} 

\IEEEpeerreviewmaketitle

\vspace{-7pt}
\section{Introduction}
Due to the growing demand in data traffic, large improvements in the spectral efficiency are required \cite{cisco}. Network densification has been identified as a possible way to achieve the desired spectral efficiency gains \cite{ghosh,alexiou}. This approach consists of deploying a large number of low powered base stations (BSs) known as small cells. With the addition of small cell BSs, the overall system is known as a heterogeneous cellular network (HetNet). Co-channel deployment of small cell BSs results in high intercell interference if their operation is not coordinated \cite{kosta}.

Interference coordination techniques such as intercell interference coordination (ICIC) has been extensively studied for multi-tier HetNet scenarios \cite{lopez,boudreau}. ICIC relies on orthogonalizing time and frequency resources allocated to the macrocell and the small cell users. Orthogonalization in time is achieved by switching off the relevant subframes belonging to the macrocell thereby reducing inter-tier interference to the small cell BSs \cite{lopez,boudreau}. Orthogonalization in frequency can be achieved with fractional frequency reuse where the users in the inner part of the cells are scheduled on the same frequency resources in contrast to the users at the cell edge whom are scheduled on available orthogonal resources. Distributed and joint power control strategies for dominant interference supression in HetNets is discussed in \cite{hossain}. The performance of multiple antenna (i.e., MIMO) HetNets using the above mentioned techniques is analyzed in \cite{dhillon2} and \cite{dhillon3}. The effects of random orthogonal beamforming with maximum rate scheduling for MIMO HetNets is studied in \cite{park}. The effects of imperfect channel state information (CSI) with limited feedback MIMO is investigated in \cite{akoum} for a two-tier HetNet.

In addition to orthogonalization, interference coordination can also be achieved by means of transmit beamforming at the BSs. However, there seems to be limited literature on transmit beamforming techniques to coordinate interference in HetNets \cite{liu,hong}. Transmit beamforming techniques have been well explored in the multiuser (MU) MIMO literature to mitigate or reduce the effects of intracell interference \cite{goldsmith,peel,sadek}. Performance superiority at low signal-to-noise-ratio (SNR) of the leakage based beamforming technique compared to zero-forcing beamforming (ZFBF) is shown in \cite{sadek}. With ZFBF, complete MU intracell interference cancellation takes place if perfect CSI is present at the BS and the number of transmit antennas exceeds the total number of receive antennas. However, leakage based beamforming focuses on maximizing the desired signal-to-leakage-noise-ratio (SLNR) without any restrictions on the number of transmit antennas. The focus of this paper is on the performance gains of a two-tier HetNet with active interference coordination. Intracell and intercell interference is coordinated by deploying leakage based beamformers at the macrocell and microcell BSs. We summarize the contributions of this paper as follows:
\begin{itemize}
\item{We evaluate the performance gains of full coordination and macro-only coordination techniques relative to no coordination for two-tier HetNets. The impact of imperfect CSI on the performance of these coordination techniques is also investigated.}
\item{We demonstrate the effect of network densification with varying degrees of BS coordination on the mean per-user signal-to-interference-plus-noise-ratio (SINR) and compare the simulated mean per-user SINR results with the analytical approximations over a wide range of SNR. The mean per-user SINR decreases with an increasing microcell count. However, we show that coordination substantially reduces the rate of SINR decrease.}
\item{We show that in the absence of coordination, network densification does not provide any gain in the sum rate, whereas with coordination, a linear increase in the sum rate is observed.}
\end{itemize}
\emph{Notation:} We use the symbols $\mathbf{A}$ and $\mathbf{a}$ to denote a matrix and a vector, respectively. $\mathbf{A^{*}}$, $\mathbf{A}^{-1}$, $\mathrm{tr}\{\mathbf{A}\}$, denote the conjugate transpose, the inverse and the trace of the matrix $\mathbf{A}$, respectively. $||\cdot{}||$ and $|\cdot{}|$ stand for the vector and scalar norms, respectively. $\mathbb{E}[\cdot{}]$ denotes the statistical expectation.

\section{System Model}
\begin{figure}[!t]
\centering
\includegraphics[width=0.9\columnwidth]{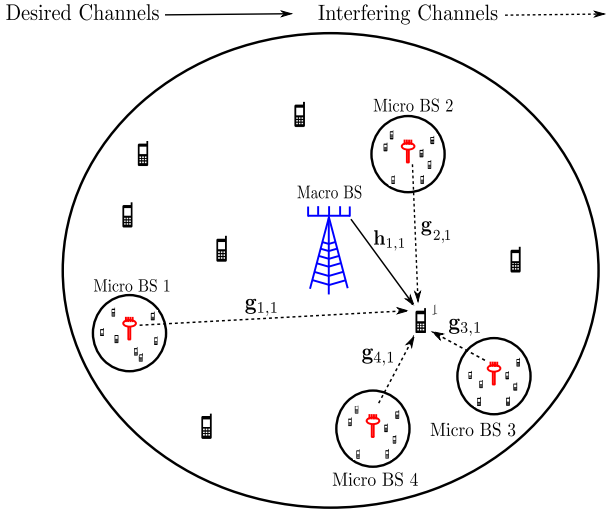}
\caption{Illustration of a two-tier HetNet with desired and interfering links to mobile user $1$ in the macrocell coverage area.}
\label{sm}
\vspace{-15pt}
\end{figure}
\subsection{System Description and Signal Model}
We consider downlink transmission in a two-tier HetNet comprising of a single macrocell BS and multiple microcell BSs, as shown in Fig.~\ref{sm}. We consider a typical scenario where the mobile users in the coverage area of a particular cell are served by the corresponding BS.  We assume that lossless and delayless backhaul links are present between each BS to exchange CSI, if desired.  We denote the total number of cells (including the macrocell and all microcells) as $N$. We denote  the number of transmit antennas on BS $n$ as $Z_{n}$ and the total number of single antenna users in cell $n$ as $k_{n}$. The received signal at mobile user $k$ in cell $n$ is given by 
\begin{flalign}
\label{rs}
&y_{n,k} =\\
&\underbrace{\mathbf{h}_{n,k}\mathbf{w}_{n,k}s_{n,k}}_{\textrm{desired signal}}\hspace{-2pt}+\hspace{-3pt} \underbrace{\sum\limits_{\substack{i=1\\i\neq{}k}}^{k_{n}}\mathbf{h}_{n,k}\mathbf{w}_{n,i}s_{n,i}}_{\textrm{intracell interference}}\hspace{-2pt}+\hspace{-2pt}\underbrace{\sum\limits_{\substack{j=1\\j\neq{}n}}^{N}\mathbf{g}_{j,k}\hspace{-2pt}\sum\limits_{q=1}^{k_{j}}\mathbf{w}_{j,q}s_{j,q}}_{\textrm{intercell interference}}\hspace{-2pt}+\hspace{-2pt}\underbrace{v_{k}}_{\textrm{noise}}.\nonumber
\end{flalign}Here, $\mathbf{h}_{n,k}$ denotes the $1\times{}Z_{n}$ complex Gaussian independent and identically distributed (i.i.d.) channel vector from the BS $n$ to user $k$. That is, $\mathbf{h}_{n,k}\sim\mathcal{CN}(0,P_{n,k})$, where $P_{n,k}$ denotes the received power from BS $n$ to user $k$. $\mathbf{w}_{n,k}$ is the $Z_{n}\times{}1$ normalized beamforming vector from BS $n$ to user $k$. $s_{n,k}$ is the desired transmitted data symbol by BS $n$ to user $k$. The transmitted data symbols are normalized such that $\mathbb{E}[|s_{n,k}|^{2}]=1$. $\mathbf{g}_{j,k}$ denotes the $1\times{}Z_{j}$ complex Gaussian i.i.d. intercell interfering channel vector from BS $j$ to user $k$ located in cell $n$. That is, $\mathbf{g}_{j,k}\sim\mathcal{CN}(0,P_{j,k})$.\footnote{Note that $\mathbf{h}$ and $\mathbf{g}$ are used to denote the desired and intercell interfering channels, respectively,  regardless of the originating BS type; i.e., $\mathbf{g}$ can represent the intercell interfering link from the macrocell BS for a particular user placed in a microcell.}$^{,}$\footnote{We drop the index $n$ from $\mathbf{g}_{j,k}$ to simplify the notation.} $v_{k}$ is the additive white Gaussian noise at receiver $k$ having an independent complex Gaussian distribution with variance $\sigma_{k}^{2}$. Finally, $P_{n,k}$ is defined as 
\begin{equation}
\label{rp}
P_{n,k}=\frac{P_{t,n}}{k_{n}}\bigg(\frac{d_{0}}{d_{n,k}}\bigg)^{\Gamma}\Phi. 
\end{equation}Here, $P_{t,n}$ refers to the total effective radiated transmit power (ERP) from BS $n$. Naturally, the ERP of the macrocell BS is higher than the microcell BSs. $d_{0}$ is a reference distance of $1$ meter (m) for far field transmit antennas, $d_{n,k}$ is the distance to mobile user $k$ from the BS $n$, $\Gamma$ is the pathloss exponent for urban macro (UMa) or urban micro (UMi) depending on the transmitting BS and $\Phi$ is the correlated shadow fading value with a standard deviation $\sigma_{\Phi}$, obtained from the Gudmundson model \cite{gudmundson} with a decorrelation distance of $10$ m. SNR with respect to BS $n$ and user $k$ is defined as $P_{t,n}/\sigma_{k}^{2}$, where $\sigma_{k}^{2}$ is the receiver noise variance at user $k$. 
\subsection{SINR Analysis with Leakage Based Beamforming}
From (\ref{rs}), the SINR at user $k$ being served by BS $n$ can be expressed as 
\begin{equation}
\label{SINR}
\gamma_{n,k}=\frac{\big|\mathbf{h}_{n,k}\mathbf{w}_{n,k}\big|^{2}}{\sigma_{k}^{2}+\sum\limits_{\substack{i=1\\i\neq{}k}}^{k_{n}}\big|\mathbf{h}_{n,k}\mathbf{w}_{n,i}\big|^{2}+\sum\limits_{\substack{j=1\\j\neq{}n}}^{N}\sum\limits_{q=1}^{k_{j}}\big|\mathbf{g}_{j,k}\mathbf{w}_{j,q}\big|^{2}}.
\end{equation}
The leakage based technique to generate beamforming vectors is as described in \cite{sadek}, where the main idea is to maximize the desired signal power relative to the noise and total interference powers caused to other users (leakage power). The SLNR for user $k$ served by the BS $n$ is defined as 
\begin{equation}
\label{SLNR}
\mathrm{SLNR}_{n,k}=\frac{\big|\mathbf{h}_{n,k}\mathbf{w}_{n,k}\big|^{2}}{\sigma_{k}^{2}+\sum\limits_{\substack{i=1\\i\neq{}k}}^{k_{n}}\big|\mathbf{h}_{n,i}\mathbf{w}_{n,k}\big|^{2}+\sum\limits_{\substack{j=1\\j\neq{}n}}^{N}\sum\limits_{q=1}^{k_{j}}\big|\mathbf{g}_{n,q}\mathbf{w}_{n,k}\big|^{2}}.
\end{equation}
For single-stream transmission (where each user is equipped with a single receive antenna), the leakage based beamforming vector desired for user $k$ being served by BS $n$ is given by the normalized version of the 
\begin{equation}
\label{SLNR-BF}
\mathbf{w}_{n,k}=(\mathbf{\tilde{H}}_{n,k}^{*}\mathbf{\tilde{H}}_{n,k}+\sigma{}_{k}^{2}\mathbf{I})^{-1}\mathbf{h}_{n,k}^{*},
\end{equation}
such that $||\mathbf{w}_{n,k}||^{2}=1$. The structure of (\ref{SLNR-BF}) remains unchanged regardless of the coordination strategy. However, the composition of $\mathbf{\tilde{H}}_{n,k}$ depends on the coordination strategy considered, as described in Section III. For the simple case of no coordination 
\begin{equation}
\label{concchannel}
\mathbf{\tilde{H}}_{n,k}=[\mathbf{h}_{n,1},\dots,\mathbf{h}_{n,k-1},\mathbf{h}_{n,k+1},\dots,\mathbf{h}_{n,k_{n}}]
\end{equation}
is the concatenated channel of all users being served by BS $n$ apart from user $k$. Assuming the distribution of intracell and intercell interference terms in (\ref{SINR}) is identical to the distribution of noise, the mean sum rate for cell $n$ can be expressed as 
\begin{equation}
\label{SR1}
\mathbb{E}[\mathrm{R}_{n}]=\mathbb{E}\Bigg[\sum\limits_{k=1}^{k_{n}}\log_{2}(1+\gamma_{n,k})\Bigg]. 
\end{equation}
The mean sum rate over $N$ cells can then be expressed as 
\begin{equation}
\label{SR2}
\mathbb{E}[\mathrm{R}_{N}]=\mathbb{E}\Bigg[\sum\limits_{\substack{j=1}}^{N}\sum\limits_{k=1}^{k_{j}}\log_{2}(1+\gamma_{j,k})\Bigg]. 
\end{equation}
From (\ref{SINR}), the mean per-user SINR can be expressed as $\mathbb{E}[\gamma_{n,k}]$. Exact evaluation of $\mathbb{E}[\gamma_{n,k}]$ is extremely cumbersome. Instead, we consider an approximation motivated by the work in \cite{yu}, which allows us to express the mean per-user SINR as 
\begin{flalign}
\label{MSINR}
\nonumber
&\mathbb{E}[{\gamma}_{n,k}]\approx\\ 
&\frac{\mathbb{E}\bigg[\big|\mathbf{h}_{n,k}\mathbf{w}_{n,k}\big|^{2}\bigg]}{\sigma_{k}^{2}+\mathbb{E}\bigg[\sum\limits_{\substack{i=1\\i\neq{}k}}^{k_{n}}\big|\mathbf{h}_{n,k}\mathbf{w}_{n,i}\big|^{2}\bigg]+\mathbb{E}\bigg[\sum\limits_{\substack{j=1\\j\neq{}n}}^{N}\sum\limits_{q=1}^{k_{j}}\big|\mathbf{g}_{j,k}\mathbf{w}_{j,q}\big|^{2}\bigg]}.
\end{flalign}The statistical expectations in both the numerator and the denominator of (\ref{MSINR}) can be evaluated further. An approach to derive the closed-form approximation of (\ref{MSINR}) is presented in the Appendix. On the other hand, (\ref{MSINR}) can be rewritten in its equivalent trace form as 
\begin{flalign}
\label{MSINR2}
\nonumber
&\mathbb{E}[\gamma_{n,k}]\approx\\
&\frac{\mathbb{E}[\mathrm{tr}\{\mathbf{w}_{n,k}^{*}\mathbf{A}\mathbf{w}_{n,k}\}]}{\sigma_{k}^{2}+\sum\limits_{\substack{i=1\\i\neq{}k}}^{k_{n}}\mathbb{E}[\mathrm{tr}\{\mathbf{w}_{n,i}^{*}\mathbf{A}\mathbf{w}_{n,i}\}]+\sum\limits_{\substack{j=1\\j\neq{}n}}^{N}\sum\limits_{q=1}^{k_{j}}\mathbb{E}
[\mathrm{tr}\{\mathbf{w}_{j,q}^{*}\mathbf{B}\mathbf{w}_{j,q}\}]},
\end{flalign}where $\mathbf{A}=\mathbf{h}_{n,k}^{*}\mathbf{h}_{n,k}$ and $\mathbf{B}=\mathbf{g}_{j,k}^{*}\mathbf{g}_{j,k}$. The expression in (\ref{MSINR2}) is used to approximate the mean per-cell sum rate over a wide range of SNR and the mean per-user SINR over a large number of channel realizations as specified in Section IV.
\subsection{Imperfect CSI Model}
It is idealistic to assume perfect CSI at all times to generate the leakage based beamforming vectors. Thus, we consider channel imperfections via channel estimation errors as mentioned in \cite{suraweera}. The imperfect channel at BS $n$ of user $k$ after introducing channel estimation errors is given by 
\begin{align}
&\hat{\mathbf{h}}_{n,k}=\rho\mathbf{h}_{n,k}+\sqrt{1-\rho^{2}}\mathbf{\Xi}_{n,k}.
\end{align}
Here, $0\leq\rho\leq1$ controls the level of CSI imperfection. $\rho=1$ results in perfect CSI and $\rho=0$ models complete uncertainty. $\mathbf{\Xi}$ is a $1\times{}Z_{n}$ complex Gaussian error vector with a statistically identical structure to $\mathbf{h}_{n,k}$. It is shown in \cite{ahn} that $\rho$ can be used to determine the impact of several factors on imperfect CSI and can be a function of the length of the estimation pilot sequence, Doppler frequency and SNR. The concatenated channel and the leakage based beamforming vector for user $k$ in cell $n$ can be expressed as (\ref{SLNR-BF}) and (\ref{concchannel}) when replacing $\mathbf{\tilde{H}}_{n,k}$ with $\mathbf{\hat{\tilde{H}}}_{n,k}$ and $\mathbf{h}_{n,k}$ with $\mathbf{\hat{h}}_{n,k}$, respectively. The SINR with imperfect CSI can be expressed as in (\ref{SINR}) when replacing $\mathbf{w}_{n,k}$ with $\hat{\mathbf{w}}_{n,k}$, $\mathbf{w}_{n,i}$ with $\hat{\mathbf{w}}_{n,i}$ and $\mathbf{w}_{j,q}$ with $\hat{\mathbf{w}}_{j,q}$, respectively. As the leakage based beamforming vectors are designed with imperfect CSI, the SINR expressed in (\ref{SINR}) will contain channel estimation errors.

\section{Two-tier Coordination Strategies} 
\begin{table}[!t]
\renewcommand{\arraystretch}{1.2}
\caption{BS Coordination Strategies and the Associated $\mathbf{\tilde{H}_{n,k}}$}
\centering
\resizebox{268pt}{43pt}{
\begin{tabular}{cc}
\bfseries Coordination Strategy  & $\mathbf{\tilde{H}}_{n,k}$ \\ \hline
No coord. (baseline case) & $[\mathbf{h}_{n,1},\dots,\mathbf{h}_{n,k-1},\mathbf{h}_{n,k+1},\dots,\mathbf{h}_{n,k_{n}}]$ \\ \hline
\multirow{2}{*}{Full coord.} & $[\mathbf{h}_{n,1},\dots{},\mathbf{h}_{n,k-1},\mathbf{h}_{n,k+1},\dots{},\mathbf{h}_{n,k_{n}},\mathbf{G}_{n,1},\dots,\mathbf{G}_{n,N}]$\\ & $\mathbf{G}_{n,m}=[\mathbf{g}_{m,1},\mathbf{g}_{m,2},\mathbf{g}_{m,3}\dots{},\mathbf{g}_{m,k_{m}}].$ \\ \hline
\multirow{2}{*} {Macro-only coord.} & If $n$ is a macro BS: equivalent to Full Coord.\\ & If $n$ is a {micro BS}: equivalent to No Coord.\\ \hline
\multirow{3}{*} {No inter-tier int.} & If $n$ is a macro BS: equivalent to No Coord.\\ & If $n$ is a micro BS: \\ & $[\mathbf{h}_{n,1},\dots{},\mathbf{h}_{n,k-1},\mathbf{h}_{n,k+1},\dots{},\mathbf{h}_{n,k_{n}},\mathbf{G}_{n,1},\dots,\mathbf{G}_{n,N-1}]$ \\
\end{tabular}}
\label{tab:tab1}
\end{table}​
In this section, we describe the BS coordination strategies considered.
\begin{itemize}
\item{\emph{No Coordination} -- In this case, each BS coordinates the desired and intracell interfering links locally. That is, the BSs only consider maximizing the SLNR of users belonging to its own coverage area. The concatenated channel used to compute the leakage based beamforming vector weights for user $k$ in cell $n$ is given in (\ref{concchannel}). We treat this strategy as the baseline case.}
\item{\emph{Full Coordination} -- In this case, we assume that each BS has knowledge of its own users desired channels and all intracell and intercell interfering channels. The channel information may be exchanged by exploiting the intercell orthogonal reference signals via the backhaul interface \cite{lee}. With the use of the fully acquired CSI for each desired and interfering link, downlink leakage based beamformers can be designed to minimize the leakage power within the cell as well as to the other cells. The concatenated channel used to compute the leakage based beamforming vector weights for user $k$ in cell $n$ can be expressed as 
\begin{flalign}
\label{FCConc}
\nonumber
\mathbf{\tilde{H}}_{n,k}=[\mathbf{h}_{n,1},\dots{},\mathbf{h}_{n,k-1},\mathbf{h}_{n,k+1},\dots{},\mathbf{h}_{n,k_{n}},\\\mathbf{G}_{n,1},\dots,\mathbf{G}_{n,N}].
\end{flalign}
Here $\mathbf{G}_{n,m}$ denotes the concatenated intercell interfering channels transmitted from BS $n$ to all users in cell $m$, given by 
\begin{flalign}
\label{ICIconc}
\mathbf{G}_{n,m}=[\mathbf{g}_{m,1},\mathbf{g}_{m,2},\mathbf{g}_{m,3}\dots{},\mathbf{g}_{m,k_{m}}].
\end{flalign}}
\item{\emph{Macro-Only Coordination} --  In this case, we assume that the macrocell BS has knowledge of the intercell interfering channels from itself to all microcell users. The macrocell BS uses this information to coordinate transmission to its own users, as well as to the users located in each microcell, respectively. The concatenated channel used to compute the leakage based beamforming weight vectors for user $k$ in cell $n$ can be expressed as (\ref{FCConc}) and (\ref{concchannel}) if $n$ is the macrocell and microcell BS, respectively.}
\item{\emph{No Inter-tier Interference} -- This is an ideal case, where we assume that no cross-tier interference exists. This means that users in a particular tier only experience intra-tier interference. Coordination is however present within each cell regardless of the tier. In computing the leakage based beamforming weight vector for user $k$ in cell $n$, the concatenated channel will be given by (\ref{concchannel}) if BS $n$ is the macrocell BS. Otherwise, for a microcell BS it is given as 
\begin{flalign}
\label{NITIConc}
\nonumber
\mathbf{\tilde{H}}_{n,k}=[\mathbf{h}_{n,1},\dots{},\mathbf{h}_{n,k-1},\mathbf{h}_{n,k+1},\dots{},\mathbf{h}_{n,k_{n}},\\\mathbf{G}_{n,1},\dots,\mathbf{G}_{n,N-1}], 
\end{flalign}} 
where $1,\dots,N-1$ refer to microcell BS indices. 
\end{itemize}
Table \ref{tab:tab1} summarizes the different BS coordination strategies with the respective structures for $\mathbf{\tilde{H}}_{n,k}$. 

\section{Simulation Results}
We consider a two-tier HetNet system comprising of a single macrocell and two microcells (unless otherwise stated). We carry out Monte-Carlo simulations to evaluate the system performance over $10000$ channel realizations. The location of the macrocell BS was fixed at the origin of the circular coverage area with radius $d_{\mathrm{M}}$. The locations of the microcell BSs inside the macrocell coverage area were uniformly generated subject to a spacing constraint. The minimum distance between two microcells was fixed to twice the radius of the microcell, i.e., $2d_{\mathrm{m}}$, such that there is no overlap between successive microcells. In Table \ref{tab:tab2}, we specify the remainder of the simulation parameters and their corresponding values.
\begin{table}[!t]
\renewcommand{\arraystretch}{1.2}
\caption{Simulation Parameters and Values}
\centering
\begin{tabular}{cc}
\bfseries Simulation Parameter  & \bfseries Value \\ \hline
Transmit antennas (Macro BS, Micro BS) & $4$, $2$  \\ \hline
BS ERP (Macro BS, Micro BS) & $46$ dBm, $30$ dBm \\ \hline
Single antenna users (Macro BS, Micro BS) & $6$, $4$\\ \hline
Cell radius ($d_{\mathrm{M}}$, $d_{\mathrm{m}})$ &$1$ km, $70$ m\\  \hline
Pathloss exponent (UMa, UMi) & $4$, $3.5$\\ \hline
Shadowing standard deviation $(\sigma_{\Phi})$ & $8$ dB
\end{tabular}
\label{tab:tab2}
\end{table}​
\begin{figure}[!t]
\centering
\includegraphics[width=0.9\columnwidth]{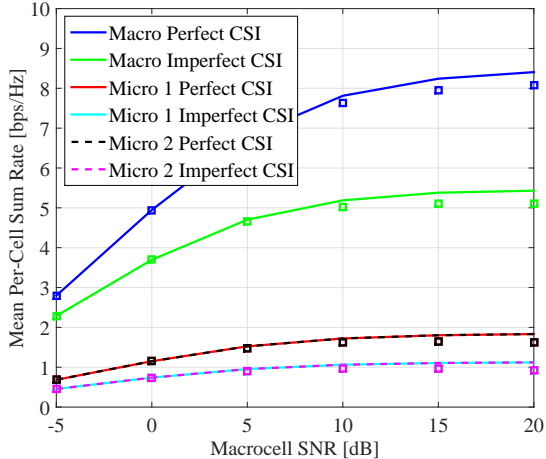}
\vspace{-5pt}
\caption{No coordination (baseline case) mean per-cell sum rate [b/s/Hz] from (\ref{SR1}) vs. macrocell SNR [dB] for perfect and imperfect CSI where $\rho=0.9$. The squares denote the approximated mean per-cell sum rates computed with (\ref{MSINR2}).}
\label{nc}
\vspace{-13pt}
\end{figure}
\begin{figure}[ht]
\centering
\includegraphics[width=0.9\columnwidth]{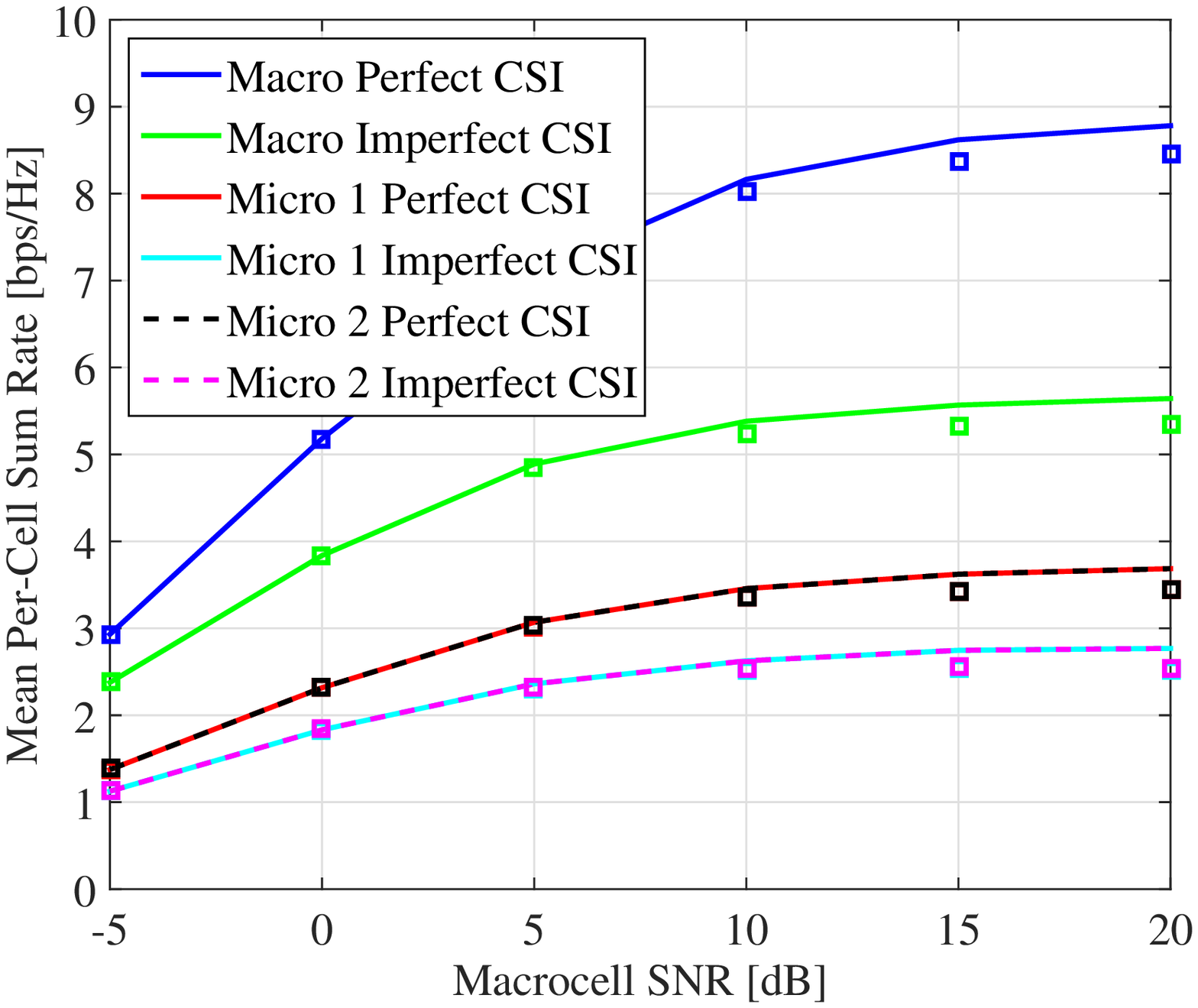}
\vspace{-5pt}
\caption{Full coordination mean per-cell sum rate [b/s/Hz] from (\ref{SR1}) vs. macrocell SNR [dB] for perfect and imperfect CSI where $\rho=0.9$. The squares denote the approximated  mean per-cell sum rates computed with (\ref{MSINR2}).}
\label{fc}
\vspace{-13pt}
\end{figure}
\begin{figure}[ht]
\centering
\includegraphics[width=0.9\columnwidth]{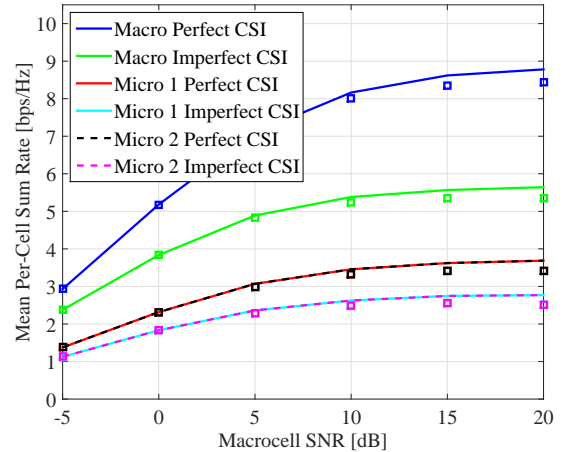}
\vspace{-5pt}
\caption{Macro-only coordinated network mean per-cell sum rate [b/s/Hz] from (\ref{SR1}) vs. macrocell SNR [dB] for perfect and imperfect CSI where $\rho=0.9$. The squares denote the approximated mean per-cell sum rates computed with (\ref{MSINR2}).}
\label{moc}
\vspace{-13pt}
\end{figure}
\begin{figure}[ht]
\centering
\includegraphics[width=0.899\columnwidth]{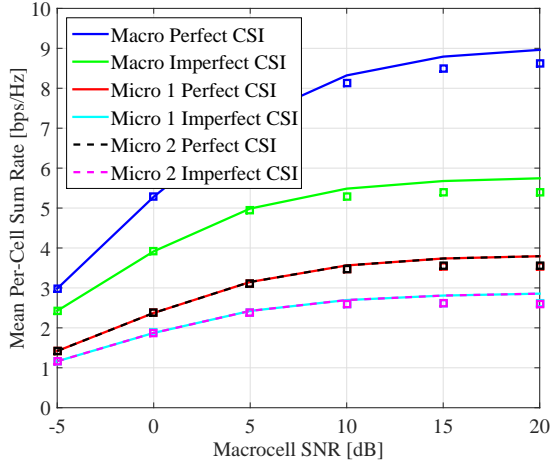}
\vspace{-11pt}
\caption{No inter-tier interference mean per-cell sum rate [b/s/Hz] from (\ref{SR1}) vs. macrocell SNR [dB] for perfect and imperfect CSI where $\rho=0.9$. The squares denote the approximated  mean per-cell sum rates computed with (\ref{MSINR2}).}
\label{niti}
\vspace{-13pt}
\end{figure}
\begin{figure}[ht]
\centering
\includegraphics[width=0.9\columnwidth]{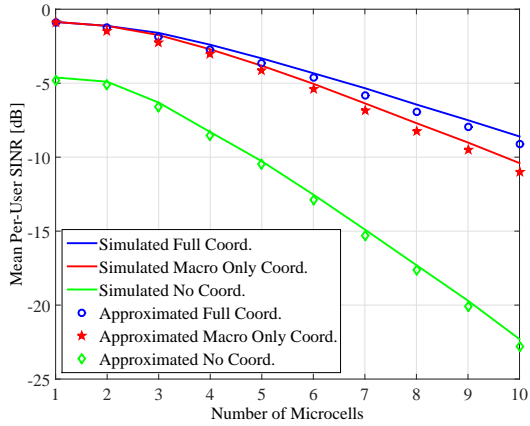}
\vspace{-5pt}
\caption{Mean per-user SINR [dB] performance vs. number of microcells at SNR=10 dB with perfect CSI for full, no and macro only coordination strategies. The approximated mean per-user SINRs are computed with (\ref{MSINR2}).}
\label{mpusinr}
\vspace{-13pt}
\end{figure}
\begin{figure}[ht]
\centering
\vspace{-5pt}
\includegraphics[width=0.9399\columnwidth]{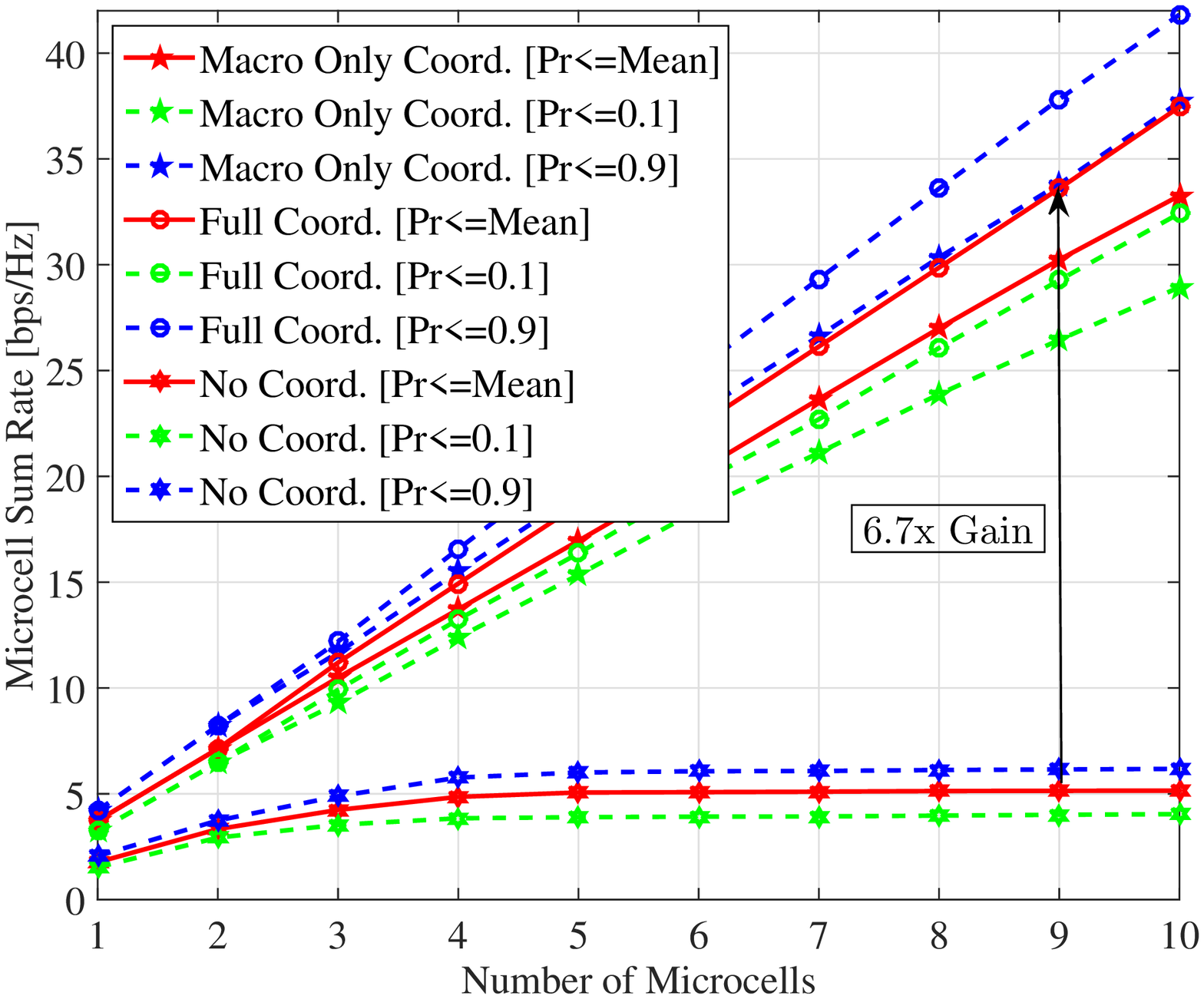}
\vspace{-11pt}
\caption{Microcell sum rate [b/s/Hz] performance from (\ref{SR2}) vs. number of microcells at SNR=10 dB with perfect CSI for full, no and macro only coordination strategies.}
\label{mrp}
\end{figure}
\begin{figure}[ht]
\centering
\vspace{-14pt}
\includegraphics[width=0.93\columnwidth]{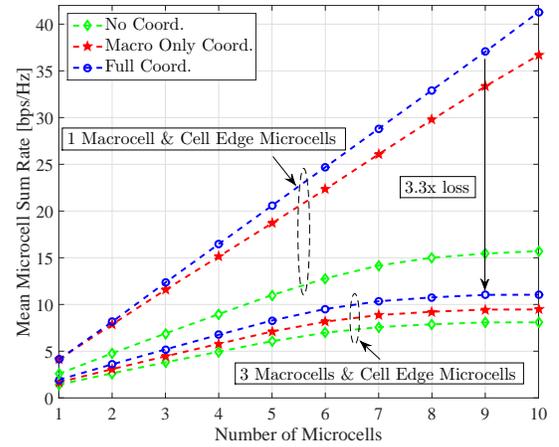}
\vspace{-11pt}
\caption{Mean microcell sum rate [b/s/Hz] per macrocell from (\ref{SR2}) vs. number of microcells at SNR=10 dB with perfect CSI for full, no and macro only coordination strategies.} 
\label{mrp2}
\vspace{-15pt}
\end{figure}
\subsection{No Coordination - Baseline Case}
Fig.~\ref{nc} shows the mean per-cell sum rate performance given by (\ref{SR1}) vs. macrocell SNR with no coordination in the HetNet. We consider perfect and imperfect CSI at the BSs. In the high SNR regime, inter-tier interference causes the mean sum rates to saturate for macrocell and microcells, respectively. The dominant factor contributing to the poor mean sum rate performance of microcell users is the large inter-tier interference from the macro BS resulting from its high transmit power. This behaviour is a result of the uncoordinated nature of the HetNet. With imperfect CSI, we again consider the mean sum rate performance with $\rho=0.9$, where further degradation in the macrocell and microcell rates can be observed. The approximated mean per-cell sum rates based on (\ref{MSINR2}) are shown to closely match the simulated responses. The variation between the simulated and analytical SINR responses can be justified from the fact that the approximation in (\ref{MSINR2}) becomes less tight with increasing SNR.
\subsection{Full Coordination}
The uncoordinated network performance can be compared to the case where the HetNet is fully coordinated. Fig.~\ref{fc} demonstrates the mean per-cell sum rate performance given by (\ref{SR1}) vs. macrocell SNR for perfect and imperfect CSI. Two major trends can be observed from the result.  First is the near $100\%$ increase in the microcell rates over the entire SNR range relative to the baseline case (Fig.~\ref{nc}). Secondly, microcell to microcell interference has a marginal impact on the macrocell user rates due to their low transmit powers. This is demonstrated by comparing Fig.~\ref{fc} to Fig.~\ref{nc}.
\subsection{Macro-Only Coordination}
As the macrocell BS is the dominant source of interference to the microcell users, we consider the case where coordination takes place at the macrocell BS only. Fig.~\ref{moc} demonstrates the mean per-cell sum rate given by (\ref{SR1}) vs. macrocell SNR performance of the macro only coordination strategy. Both the macro and microcell rates are found to be approximately equivalent to the full coordination case, observed by comparing Fig.~\ref{fc} and Fig.~\ref{moc}. This suggests that if we can coordinate the transmission to minimize the most dominant source of interference, we are able to achieve near full coordination performance. Moreover, this strategy significantly reduces the backhaul overheads by eliminating the need to equip the microcell BSs with out-of-cell CSI. 
\subsection{No Inter-tier Interference}
Fig.~\ref{niti} depicts the mean per-cell sum rate performance given by (\ref{SR1}) vs. macrocell SNR of the no inter-tier interference coordination strategy. Due to zero cross-tier interference, this strategy results in superior mean per-cell sum rate performance in comparison with the other coordination strategies. It is worth comparing Fig.~\ref{niti} to Fig.~\ref{fc}, and noting that the mean sum rate performance of full coordination approaches the performance of no inter-tier interference. This demonstrates the value of BS coordination in a HetNet. 

\subsection{Network Densification}
The effect of increasing the microcell density is shown in Fig.~\ref{mpusinr}, where we plot the mean per-user SINR as a function of the number of microcells. We observe that the mean per-user SINR decreases linearly with increasing number of microcells. When the number of microcells is less than 5, there is a marginal difference between macro only coordination and full coordination mean per-user SINR. This suggests that at low microcell density, it is advantageous to avoid paying the high price of backhaul overheads for full coordination performance. When there are more than 5 microcells, the gap between full coordination and macro-only coordination techniques starts to increase. Approximately, a $2$ dB difference in the mean per-user SINR is seen with $10$ microcells in the system. The difference in the slopes of the various strategies demonstrates the impact of BS coordination in a HetNet with network densification. Thus, coordination arrests the rate of decay of the mean per-user SINR in a HetNet. In addition to the above, the result demonstrates the validity of the approximated mean per-user SINR in (\ref{MSINR2}). These are shown to closely match the simulated mean per-user SINR performance for all the coordination techniques. Fig.~\ref{mrp} shows the microcell sum rate performance as defined in (\ref{SR2}) at the mean, $10$th and $90$th percentiles with respect to number of microcells at a SNR of $10$ dB. With full coordination, the microcell sum rate increases linearly with the number of microcells, as majority of the interference is being suppressed by the leakage based beamformers. A similar trend can be observed for the macro only coordination case, however the microcell sum rate performance gains are lower compared to the full coordination case as the number of microcells increases. The no coordination case suffers from strong macro and other microcell interference resulting in a saturated sum rate at higher number of microcells.

\subsection{Impact of Multiple Macrocells}
We now study the effect of deploying multiple macrocell BSs on the microcell sum rate performance. For comparison purposes, we consider scenarios with both single and three overlapping macrocells with inter-site distances of $1$ km. In both cases, a maximum of $10$ microcell BSs are randomly dropped at the edge of the macrocell at a radius of $877\footnote{This value is obtained numerically at the $10$th percentile of the received SNR cumulative distribution function (CDF) averaged over the macrocell user distances.}<d_{\mathrm{M}}<1000\hspace{2pt}\mathrm{m}$, such that the minimum distance between successive microcell BSs is $2d_{\mathrm{m}}$. Fig.~\ref{mrp2} shows the mean microcell sum rate as a function of the number of microcells for both the single and three macrocell BSs cases at a SNR of 10dB. It is seen that the sum rate of the single macrocell BS case is significantly higher than the three overlapping macrocell BS case. This is due to higher aggregate intercell interference resulting from other macrocells and microcells located within these macrocells. Compared to Fig.~\ref{mrp} where the microcells are randomly placed anywhere within the macrocell coverage area, the no coordination performance benefits the most from the microcells being deployed at the edge of the macrocell. This can be seen from the mean sum rate, as it shows a linear growth up to 7 microcells in comparison with 3 microcells. We also observe that the improvement in mean sum rate with cell edge deployment of microcells is higher for the no coordination strategy. At 10 microcells, the increase in the mean sum rate for the full coordination strategy is approximately 3.6 bps/Hz, while the increase with no coordination is about 10 bps/Hz.

\section{Conclusion}
In this paper, we demonstrate the rate gains provided by BS coordination in HetNets. With BS coordination, the sum rate is seen to increase linearly  and the mean per-user SINR decreases linearly with the number of microcells. However, the rate of mean per-user SINR degradation is reduced significantly with increased degrees coordination at the BSs in the HetNet. At a low density of microcells, macro-only coordination performs close to full coordination. However, this is not the case with a higher density of microcells where increasing amounts of interference from the microcells is being added. In addition to the above, the impact of multiple macrocells is also investigated. Here, degradation in the mean microcell sum rate is observed for all the respective coordination strategies in comparison to the case where only one macrocell is present.

\appendix
The numerator of (\ref{MSINR}) can be further evaluated as shown below. Substituting the definition of $\mathbf{w}_{n,k}$ gives
\begin{flalign}
\label{app1}
&\mathbb{E}\bigg[\big|\mathbf{h}_{n,k}\mathbf{w}_{n,k}\big|^{2}\bigg]=\mathbb{E}\bigg[\big|\mathbf{h}_{n,k}(\mathbf{\tilde{H}}_{n,k}^{*}\mathbf{\tilde{H}}_{n,k}+\sigma_{k}^{2}\mathbf{I})^{-1}\mathbf{h}_{n,k}^{*}\big|^{2}\bigg].
\end{flalign}
Using an eigenvalue decomposition, (\ref{app1}) can be rewritten as 
\begin{align}
\nonumber
\label{app2}
&\mathbb{E}\bigg[\big|\mathbf{h}_{n,k}\mathbf{w}_{n,k}\big|^{2}\bigg]=\mathbb{E}\bigg[\big|\mathbf{h}_{n,k}(\mathbf{X\Lambda{}X}^{*}+\sigma_{k}^{2}\mathbf{I})^{-1}\mathbf{h}_{n,k}^{*}\big|^{2}\bigg]\\
&\hspace{68pt}=\mathbb{E}\bigg[\big|\boldsymbol{\delta}_{n,k}(\mathbf{\Lambda}+\sigma_{k}^{2}\mathbf{I})^{-1}\boldsymbol{\delta}_{n,k}^{*}\big|^{2}\bigg],
\end{align}
where $\boldsymbol{\delta}_{n,k}=\mathbf{h}_{n,k}\mathbf{X}$ has the same statistics as $\mathbf{h}_{n,k}$ as $\mathbf{X}$ is a unitary matrix. Hence, 
\begin{equation}
\label{app3}
\mathbb{E}\bigg[\big|\mathbf{h}_{n,k}\mathbf{w}_{n,k}\big|^{2}\bigg]=\mathbb{E}\bigg[\big|\sum\limits_{i=1}^{Z_{n}}(\mathbf{\Lambda}_{ii}+\sigma_{k}^{2})^{-1}|\boldsymbol{\delta}_{n,k,i}|^{2}\big|^{2}\bigg],
\end{equation}
where $\boldsymbol{\delta}_{n,k,i}$ is the $i$th element of $\boldsymbol{\delta}_{n,k}$. Since $\boldsymbol{\delta}_{n,k,i}$ is a zero mean complex Gaussian random variable with variance $P_{n,k}$, it follows that $|\boldsymbol{\delta}_{n,k,i}|^{2}$ is an exponential random variable with mean $P_{n,k}$. Using the standard properties of the exponential random variable, \eqref{app3} can be expressed as 
\begin{equation}
\label{app4}
=P_{n,k}^{2}\mathbb{E}\bigg[\bigg(\sum\limits_{i=1}^{Z_{n}}(\mathbf{\Lambda}_{ii}+\sigma_{k}^{2})^{-1}\bigg)^{2}+\sum\limits_{i=1}^{Z_{n}}(\mathbf{\Lambda}_{ii}+\sigma_{k}^{2})^{-2}\bigg].
\end{equation}
A similar approach can be taken to evaluate the mean per-user intracell and intercell interference powers. Further averaging over the eigenvalues, $\mathbf{\Lambda}_{ii}$ in (\ref{app4}) is possible as the density of eigenvalues is known. However, due to space limitations we leave this approach for future work.


\end{document}